\documentclass[oneside,english,reqno,12pt]{amsart}
\usepackage[T1]{fontenc}
\usepackage[latin1]{inputenc}
\usepackage{a4wide}

\makeatletter

\usepackage{graphicx}
\usepackage{pstricks}

\usepackage{babel}
\makeatother
\begin{document}
\selectlanguage{english}

\title{From Defects to Boundaries }

\maketitle
\begin{center}Z. Bajnok%
\footnote{bajnok@afavant.elte.hu%
} and A. George%
\footnote{ag160@garfield.elte.hu%
}\end{center}

\begin{center}$^{1}$\textit{HAS, Theoretical Physics Research Group,}\\
 \textit{Institute for Theoretical Physics, E\"{o}tv\"{o}s University}\\
 \textit{H-1117 Budapest, P\'{a}zm\'{a}ny s. 1/A, Hungary.}\end{center}

\begin{center}$^{2}$\textit{Department of Mathematics, The University
of York}\\
 \textit{Heslington, York, YO10 5DD, England.}\end{center}

\begin{abstract}
In this paper we describe how relativistic field theories containing
defects are equivalent to a class of boundary field theories. As a
consequence previously derived results for boundaries can be directly
applied to defects, these results include reduction formulas, the
Coleman-Thun mechanism and Cutcosky rules. For integrable theories
the defect crossing unitarity equation can be derived and defect operator
found. For a generic purely transmitting impurity we use the boundary
bootstrap method to obtain solutions of the defect Yang-Baxter equation.
The groundstate energy on the strip with defects is also calculated. 
\end{abstract}

\section{Introduction}

Over the last decade there has been a growing theoretical and experimental
interest in systems containing defects or impurities. Although experimental
investigations, such as into quantum dots, require a more general
description the theoretical focus has largely been on the integrable
aspects of defect systems. The defect Yang Baxter equations (or RT
relations), unitarity and defect crossing unitarity relations were
formulated in \cite{DMS1,DMS2}, however no derivation of defect crossing
unitarity was presented, it was simply conjectured by analogy to the
bulk theory. They showed the absence of simultaneous transmission
and reflection from an integrable defect for theories with diagonal
bulk scattering matrices, this was generalized for non-diagonal theories
with very general, even dynamical, defects in \cite{CAFG}. An analysis
of RT algebras \cite{MRS1,MRS2} and a proposal to avoid this `no
go theorem' for simultaneous transmission and reflection can be found
in \cite{R}. Recently a Lagrangian description of integrable defects
has been introduced in \cite{BCZ1,BCZ2}, including a discussion on
how the boundary Lax pair approach may be adapted for defect systems.
For a quantum mechanical treatment of the defect problem we refer
to \cite{ChFT}, while for higher dimensional conformal defects to
\cite{OFO,EGK,OM}. 

In this paper, by extending the `folding trick' (formulated previously
for free \cite{OA,Sa} or conformal \cite{EGK,QS} theories), we show
that any relativistic defect theory, integrable or not, can be described
as a certain type of boundary theory. As a consequence previous results
developed for boundary theories, such as perturbation theory, reduction
formulas, Coleman-Thun mechanism and Cutkosky rules \cite{BBT1,BBT2},
can be directly applied to defects. 

For simplicity we restrict the scope of this letter to real relativistic
scalar fields in $1+1$ dimensions in order not to complicate the
notations and concentrate on the principal issues. Similarly we restrict
ourselves to strong, i.e. parity invariant, bulk theories. Generalizations
of these results, to higher dimensions, or to more fields are straightforward. 

By applying the defect-boundary correspondence for integrable theories
we can use the results from integrable boundary theories, such as
the derivation of boundary crossing unitarity \cite{GZ}, the definition
of the boundary state, boundary bootstrap, etc. for defects. Since
the defect theory and the proposed boundary theories are equivalent
we can use results from the study of defects, such as the no go theorem,
to make statements about the behaviour of these boundary theories.
We show how one can construct solutions of the boundary Yang-Baxter
equation by bootstrapping on a known solution. We apply this method
for purely transmitting impurity by applying the defect-boundary correspondence
and obtain non-trivial solutions for any bulk theory. These solutions
have a generic form and correspond to particles with imaginary rapidities
`trapped' at the defect. Finally we use the defect operator, obtained from 
the boundary state via the defect-boundary correspondence, to perform a 
thermodynamic Bethe ansatz calculation of the ground state energy of the 
system defined on the strip with defects. 

\newcommand{\PhiL}{\Phi_{\mathbb{L}}}

\newcommand{\PhiR}{\Phi_{\mathbb{R}}}

\newcommand{\PsiL}{\Psi_{\mathbb{L}}}

\newcommand{\PsiR}{\Psi_{\mathbb{R}}}

\newcommand{\TheoL}{\mathbb{L}}

\newcommand{\TheoR}{\mathbb{R}}

\newcommand{\VL}{V_{\mathbb{L}}}

\newcommand{\VR}{V_{\mathbb{R}}}

\newcommand{\WL}{W_{\mathbb{L}}}

\newcommand{\WR}{W_{\mathbb{R}}}

\newcommand{\RL}{R_{\mathbb{L}}}

\newcommand{\RR}{R_{\mathbb{R}}}

\newcommand{\TL}{T_{\mathbb{L}}}

\newcommand{\TR}{T_{\mathbb{R}}}

\newcommand{\SL}{S_{\mathbb{L}}}

\newcommand{\SR}{S_{\mathbb{R}}}

\newcommand{\inL}{|in\rangle_{\mathbb{L}}}

\newcommand{\inR}{|in\rangle_{\mathbb{R}}}

\newcommand{\outL}{|out\rangle_{\mathbb{L}}}

\newcommand{\outR}{|out\rangle_{\mathbb{R}}}

\newcommand{\inPhi}{|in\rangle_{\Phi}}

\newcommand{\inPsi}{|in\rangle_{\Psi}}

\newcommand{\outPhi}{|out\rangle_{\Phi}}

\newcommand{\outPsi}{|out\rangle_{\Psi}}

\section{Defect-boundary correspondence}

If a field, $\PhiL$, is restricted onto left half line $x\leq0$
by a boundary located at $x=0$ the bulk and boundary interactions
are described by the Lagrangian \begin{equation}
\mathcal{L}=\Theta(-x)\left(\frac{1}{2}(\partial_{t}\PhiL)^{2}-\frac{1}{2}(\partial_{x}\PhiL)^{2}-\VL(\PhiL)\right)-\delta(x)U(\PhiL)\,\,.\label{boundaryL}\end{equation}
 Here the boundary potential $U(\PhiL)$ is understood to only depend
on the field through the value of the field at the boundary $\PhiL(0,t)$.
The corresponding equation of motion and boundary condition are \begin{align}
 & {(-\partial_{t}^{2}+\partial_{x}^{2})\PhiL(x,t)=\frac{\partial\VL}{\partial\PhiL}\,\,;}\quad x\leq0\;,\quad & {\partial_{x}\PhiL(x,t)|_{x=0}=-\frac{\partial U}{\partial\PhiL}\,\,.}\label{eombcL}\end{align}
 The Dirichlet boundary condition can be described either as a limit
of the above one, or can be obtained directly from the Lagrangian
with $U=0$ but demanding $\PhiL(0,t)=\varphi$, where $\varphi$
is a constant. 

The parity transform, $\mathcal{P}:x\leftrightarrow-x$, of this theory
is another theory, which lives on the right half line $x\geq0$. We
call the original theory $\TheoL\,$ and its parity transform $\TheoR$,
which satisfies the equation of motion and boundary condition \begin{align}
 & {(-\partial_{t}^{2}+\partial_{x}^{2})\PhiR(x,t)=\frac{\partial\VR}{\partial\PhiR}\,\,;}\quad x\geq0\;,\quad & {\partial_{x}\PhiR(x,t)|_{x=0}=\frac{\partial U}{\partial\PhiR}\,\,.}\label{eombcR}\end{align}
 The two theories are equivalent: $\VR=\VL$ and the solutions of
the two theories are connected by the parity transformation. Similarly
the parity transform of the $\TheoR$ theory is the $\TheoL$ theory.
Observe that this is not a symmetry of one theory but merely two different
descriptions of the same theory. 

The quantum version of these theories are also equivalent, which can
be seen as follows. The quantum theories are defined following \cite{BBT1}:
The interaction is supposed to switched on adiabatically in the remote
past and switched off in the remote future. The asymptotic Hilbert
spaces are then the free Hilbert spaces and, being equivalent, are
related by a unitary transformation called the $R$-matrix. This matrix
is connected to the Greens functions via the boundary reduction formula,
and can be computed in Lagrangian perturbation theory. 

Now let us see how the identification of the two theories goes: The
$\TheoL$ theory has asymptotic `\emph{in}' states, $\inL$, containing
$n_{i}$ free particles with real momentum $k_{i}>0$, (i.e. traveling
toward the boundary), located at $x=-\infty$ in the order of decreasing
momentum, highest momentum furthest to the left. The asymptotic `\emph{out}'
states, $\outL$, are composed of $n_{f}$ free particles with real
momentum $k_{f}<0$, i.e. traveling away form the boundary, also located
at $x=-\infty$ in the order of increasing momentum, most negative
furthest to the left. In the $\TheoR$ theory the `\emph{in}' states,
$\inR$, contain particles with $k_{i}<0$, while `\emph{out}' states,
$\outR$, with momentum $k_{f}>0$, and both the particles of the
`\emph{in}' and `\emph{out}' states are located at $x=\infty$ and
ordered such that the fastest moving particles are furthest to the
right. Clearly the particles of both theories behave in the same way,
first traveling towards the boundary, then, after reflecting, away
from the boundary. The states of $\TheoL$ and those of $\TheoR$
are equivalent and are mapped to each other by the parity operator
$\mathcal{P}$, which does nothing but changes the sign of the particles
momenta. 

The $R$-matrix is a unitary operator which connects the asymptotic
`\emph{in}' states to the `\emph{out}' states of each theories. They
are written as $\RL$ and $\RR$ respectively for $\TheoL$ and $\TheoR$
and both can be computed from the corresponding boundary reduction
formulae. These formulae contain the correlation functions of the
fields $\PhiL$ and $\PhiR$ and can be computed using perturbation
theory derived from \eqref{eombcL} and \eqref{eombcR}, respectively.
Any term in the two perturbative expansions -- and such a way in the
correlation functions -- are related by the parity operator, $\mathcal{P}$,
moreover the reduction formulae are the parity transformed of each
other. As a consequence the $R$-matrices are related by parity, $\mathcal{P}:\RL\leftrightarrow\RR$.
Thus $\TheoL\,$ and $\TheoR\:$ are equivalent theories on the quantum
level. 

Defect theories are simply an $\TheoL\,$ and a (possibly different)
$\TheoR\,$ type theory connected at the origin by the defect, they
can be described by the Lagrangian \begin{align}
\mathcal{L}= & \,\,\Theta(-x)\left(\frac{1}{2}(\partial_{t}\PhiL)^{2}-\frac{1}{2}(\partial_{x}\PhiL)^{2}-\VL(\PhiL)\right)\nonumber \\
+ & \,\,\Theta(x)\left(\frac{1}{2}(\partial_{t}\PsiR)^{2}-\frac{1}{2}(\partial_{x}\PsiR)^{2}-\WR(\PsiR)\right)-\delta(x)U(\PhiL,\PsiR)\,\,.\label{defectL}\end{align}
 If the fields $\PhiL$ and $\PsiR$ are not the same then the equation
of motion and boundary conditions for $\PhiL$ are given by \eqref{eombcL}
and those for $\PsiR$ by \eqref{eombcR} with $\PsiR$ replacing
$\PhiR$. Note that, for a general defect, there is no requirement that the
values of the two fields at the defect be the same. 
If the two fields are in fact the same we call
the defect system an `impurity' system. It is described by the Lagrangian
\begin{equation}
\mathcal{L}=\left(\frac{1}{2}(\partial_{t}\phi)^{2}-\frac{1}{2}(\partial_{x}\phi)^{2}-V(\phi)\right)-\delta(x)U(\phi)\,\,,\label{impurityL}\end{equation}
 where the field $\phi$ exists on the whole line. This system has
the usual equation of motion and an `impurity condition' \begin{align}
\lim_{\epsilon\rightarrow0}\left(\partial_{x}\phi(x,t)|_{x=\epsilon}-\partial_{x}\phi(x,t)|_{x=-\epsilon}\right)=\frac{\partial U}{\partial\phi}\,\,.\label{ic}\end{align}
 Note that the left hand side of \eqref{ic} does not reduce to zero
as solutions for these fields generally have discontinuities in the
gradient at the impurity. The impurity Lagrangian \eqref{impurityL}
can be written in the form of a defect Lagrangian \eqref{defectL}
if we write $\phi=\PhiL$, $V(\phi)=\VL(\PhiL)$ for $x\leq0$ and
$\phi=\PsiR$, $V(\phi)=\WL(\PsiR)$ for $x\geq0$, and impose the
requirement that ${\PhiL|_{x=0}=\PsiR|_{x=0}=:\phi_{0}}$. The boundary
conditions for the fields of the defect Lagrangian now becomes equivalent
to the impurity condition \begin{equation}
\partial_{x}\PsiR|_{x=0}-\partial_{x}\PhiL|_{x=0}=\frac{\partial U}{\partial\phi_{0}}\,\,.\label{ibc}\end{equation}
 We will generally assume that we are dealing with a defect system,
with the understanding that impurities can be dealt with in a similar
fashion by applying the impurity conditions instead of defect conditions.
When there are significant differences between how the impurity and
defect theories need approaching we will explicitly discuss each separately. 

By making a parity transformation on the $\TheoR$ theory, $\PsiR\rightarrow\PsiL$,
we can equivalently describe the defect theory as \begin{align}
\mathcal{L}= & \,\,\Theta(-x)\left(\frac{1}{2}(\partial_{t}\PhiL)^{2}-\frac{1}{2}(\partial_{x}\PhiL)^{2}-\VL(\PhiL)\right)\nonumber \\
+ & \,\,\Theta(-x)\left(\frac{1}{2}(\partial_{t}\PsiL)^{2}-\frac{1}{2}(\partial_{x}\PsiL)^{2}-\WL(\PsiL)\right)-\delta(x)U(\PhiL,\PsiL)\,\,.\label{boundaryL2}\end{align}
 The equations of motion and boundary condition for $\PsiL$ and $\PsiR$
are related like \eqref{eombcL} and \eqref{eombcR}. The equation
of motion and boundary condition for $\PhiL$ remain unchanged. For
an impurity theory the same transformation can be made after it is
written as a defect theory as described above. The boundary conditions
become \begin{align}
 & {\partial_{x}\PsiL|_{x=0}+\partial_{x}\PhiL|_{x=0}=-\frac{\partial U}{\partial\phi_{0}}\,\,;} & {\PhiL|_{x=0}=\PsiL|_{x=0}=:\phi_{0}\,\,.}\label{ibc2}\end{align}
 The Lagrangian \eqref{boundaryL2} describes a boundary field theory
with two self interacting fields, each restricted to the left half
line, these fields only interact with each other at the boundary.
The two theories \eqref{defectL} and \eqref{boundaryL2} are classically
equivalent, the solutions can be mapped to each other by acting on
one field with the parity operator ${\mathcal{P}:\PsiL\leftrightarrow\PsiR}$. 

They are also equivalent at the quantum level, which can be seen similarly
to the previous case by defining them in the framework of \cite{BBT1}.
The `\emph{in}' states in the defect theory consist of $\TheoL$ type
particles, belonging to field $\PhiL$, with $k_{i}>0$ approaching
the defect from the left and $\TheoR$ particles, belonging to field
$\PsiR$, with $k_{i}<0$ approaching from the right, as before the
order of these particles is such that the fastest moving particles
are furthest from the defect. The `\emph{in}' state can be written
as $\inL\oplus\inR$. In the `\emph{out}' state we have $\TheoL$
and $\TheoR$ particles traveling away from the defect with $k_{f}<0$
and $k_{f}>0$, and each on their respective sides of the defect,
and ordered as before. This can be written as $\outL\oplus\outR$. 

There exists a unitary operator which connects the `\emph{out}' and
`\emph{in}' asymptotic states, called the $RT$-matrix. 
The ($\TheoL\TheoL$)
element of this matrix is the reflection matrix $\RL$ which gives
the amplitudes for process connecting $\inL$ with $\outL$. Similarly
the ($\TheoR\TheoR$) element is the reflection matrix $\RR$ which
gives the amplitudes of processes connecting $\inR $ with $\outR$. 
The ($\TheoL\TheoR$) element is the transition matrix $\TL$ 
connecting $\inL$ with $\outR$ and the ($\TheoR\TheoL$) element is the
transition matrix $\TR$ connecting $\inR$ with $\outL$. 

The adiabatic description of \eqref{boundaryL2} has `\emph{in}' states
which contains particles with momentum $k_{i}<0$ of both types $\PhiL$
and $\PsiL$, approaching the boundary from $x=-\infty$ with the
usual ordering. These states can be written as $\inPhi\oplus\inPsi$.
The `\emph{out}' states have the same type of particles but with $k_{f}<0$,
with the same ordering and written as $\outPhi\oplus\outPsi$. We
have the usual $R$-matrix, $\mathcal{R},$ which connects the `\emph{in}'
and `\emph{out}' states. It has four elements the $(\Phi\Phi)$ and
$(\Psi\Psi)$ elements connect the $\inPhi$ states to $\outPhi$  and 
$\inPsi$ to $\outPsi$ states, respectively, and are denoted $\mathcal{R}_{\Phi\Phi}$ 
and $\mathcal{R}_{\Psi\Psi}$. The $(\Phi\Psi)$ and $(\Psi\Phi)$
elements, denoted $\mathcal{R}_{\Phi\Psi}$ and $\mathcal{R}_{\Psi\Phi}$,
respectively give the amplitudes of processes connecting $\inPhi$ with $\outPsi$ 
and $\inPsi$ to $\outPhi$. 
We can define a parity operator that acts on one of the fields of
the boundary system only, $\mathcal{P}_{\Psi}:\PsiL\leftrightarrow\PsiR$,
but leaves the $\Phi$ field unchanged. This operator maps the asymptotic
states of the defect and of the boundary theory to each other. To
see how the $\mathcal{R}$ and $RT$ matrices are related one has
to relate them to the corresponding correlation functions via the
reduction formula, which finally has to be computed from the Lagrangian.
Analyzing the action of the parity operator at each step we obtain
the following identifications:\begin{equation}
\mathcal{P}_{\Psi}:\mathcal{R=}\left(\begin{array}{cc}
\mathcal{R}_{\Phi\Phi} & \mathcal{R}_{\Psi\Phi}\\
\mathcal{R}_{\Phi\Psi} & \mathcal{R}_{\Psi\Psi}\end{array}\right)\leftrightarrow RT=\left(\begin{array}{cc}
\RL & \TR\\
\TL & \RR\end{array}\right)\label{RRT}\end{equation}

Now we can exploit this correspondence. All that we have learned already
for boundary theories can be applied straightforwardly to defect theories:
we can present perturbation theory, reduction formulas and in the
view of \cite{BBT2}, the derivation of Coleman-Thun mechanism and
Cutkosky rules for defect theories is also straightforward. Note that
the generalization for higher dimensions with a codimension $1$ defect
is obvious, just as is the accommodation of more fields. Also $\partial_{t}\Phi$
and $\partial_{t}\Psi$ dependent potentials or boundary conditions
that relate $\Phi,\partial_{x}\Phi$ linearly to $\Psi,\partial_{x}\Psi$
 can be incorporated. 

\section{Integrable aspects, consequences}

\smallskip
We will now focus on the integrable aspects of the correspondence
between defects and boundaries. Classical integrability in the presence
of boundaries has been the subject of several years of study. All
the methods developed for boundary integrable systems are now available
for the study of integrable defect or impurity theories. For example
Skylanin's formalism \cite{S} which allows the construction of the
conserved charges for the boundary field, and can be used to find
the integrable boundary conditions \cite{MI} can now be used on defect
or impurity systems. The Lax pair for defect systems is discussed
in \cite{BCZ1,BCZ2} and is equivalent to the boundary Lax pair approach
described in \cite{BCDR}. 

At the quantum level the existence of infinitely many conserved charges
give severe restriction on the allowed processes. As a consequence
the $R$-matrix of a boundary system can be decomposed into a product
of two particle $S$-matrices and one particle $R$-matrices. The
former entirely describes the scattering of two particles in the bulk
of the field, the latter describes the reflecting of a single particle
from the boundary. Similarly for defect systems the $RT$-matrix can
be decomposed into a product of the, possibly different, two particle
$S$-matrices of the two fields either side of the defect and one
particle $RT$-matrices describing the reflection and transmission
of individual particles through the defect. The integrable one particle
$R$-matrix and $RT$-matrix depend on the particle energy through
the absolute value of the rapidity, $\theta$, of the incident
particle. The two particle $S$-matrices depend on the absolute
value of the difference between the rapidities of the scattering
particles. The asymptotic states of the integrable boundary and defect
systems have the same description as in the non-integrable cases,
with the additional condition that, in both cases, the number of particles
in the `\emph{out}' state must be the same as the number of particles
in the `\emph{in}' state and the conserved charges of the two asymptotic
states must coincide. The integrable \emph{$\mathcal{R}$} and $RT$-matrices
are unitarity operators which map one particle `\emph{in}' states
to one particle `\emph{out}' states and are related as described in
\eqref{RRT}. 

The integrable defect theory has two $S$-matrices, $\SL$ and $\SR$,
respectively describing the two particle scattering in the $\TheoL$
and $\TheoR$ fields. The integrable boundary theory has a single
$S$-matrix describing the two particle scattering of the bulk field.
If the bulk field is composed of two noninteracting fields, $\Phi$
and $\Psi$, as in the case described as equivalent to a defect theory,
then the $S$-matrix, $\mathcal{S},$ has a block diagonal form with
diagonal elements $\mathcal{S}_{\Phi\Phi}$, $\mathcal{S}_{\Phi\Psi}$,
$\mathcal{S}_{\Psi\Phi}$ and $\mathcal{S}_{\Psi\Psi}$. The element
$\mathcal{S}_{\Phi\Phi}$ describes the scattering of two particles
both of the field $\Phi$, similarly $\mathcal{S}_{\Psi\Psi}$ describes
the scattering of two particles of field $\Psi$, $\mathcal{S}_{\Phi\Psi}$
and $\mathcal{S}_{\Psi\Phi}$ both describe the scattering between
two particles where one belongs to each of the separate fields $\Phi$
and $\Psi$. Since these $S$-matrices can be computed from the bulk
Lagrangian, (where the $\Phi$ and $\Psi$ particles interact only
with themselves) we know, that \begin{equation}
\mathcal{S}_{\Psi\Phi}=\mathcal{S}_{\Phi\Psi}=1\quad;\qquad\mathcal{S}_{\Phi\Phi}\equiv\SL\quad;\qquad\mathcal{S}_{\Psi\Psi}\equiv\SR\quad.\label{DefS}\end{equation}
(A $S$-matrix of a different form has been used in \cite{CMR1,CMR2} to
describe an integrable defect of a non-relativistic theory.) 

Factorizability of the defect system gives the defect Yang-Baxter,
or reflection transmission equations, which was described by several
authors \cite{DMS1,DMS2,CAFG,R}. Here we just present one of them
and the corresponding equation 

\begin{center} \includegraphics[
height=4cm,
keepaspectratio]{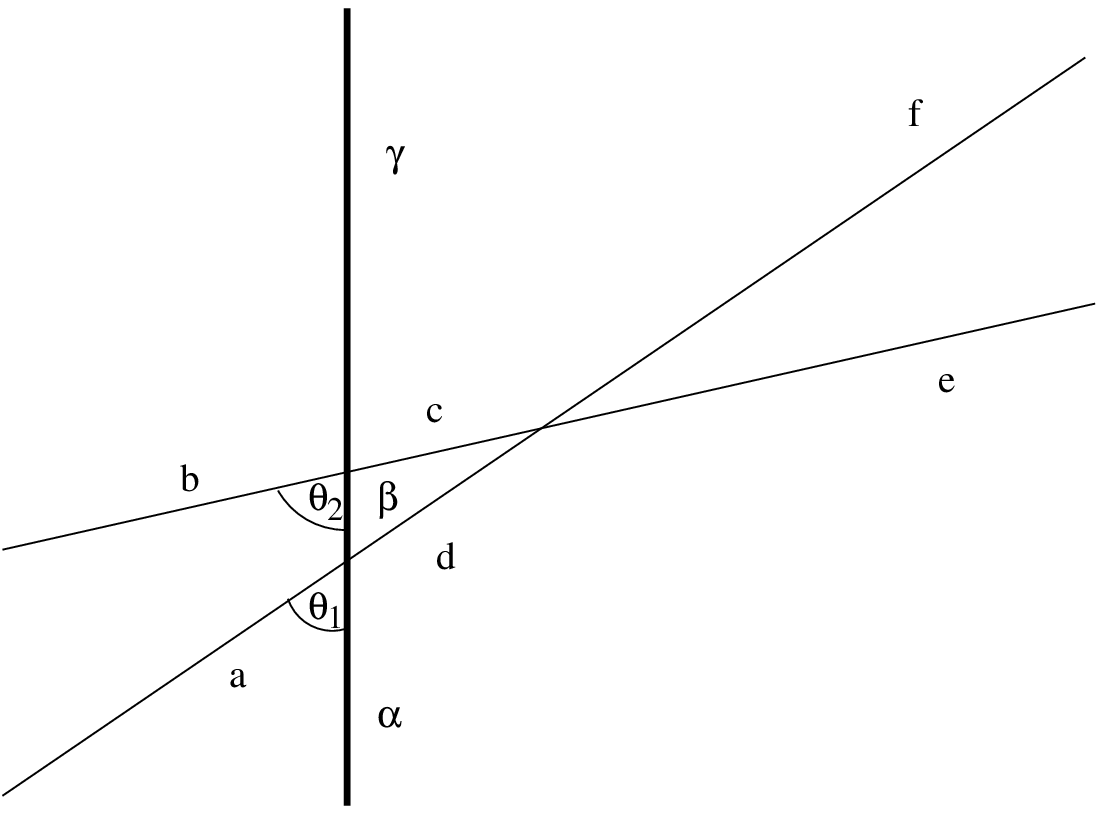} \includegraphics[
height=4cm,
keepaspectratio]{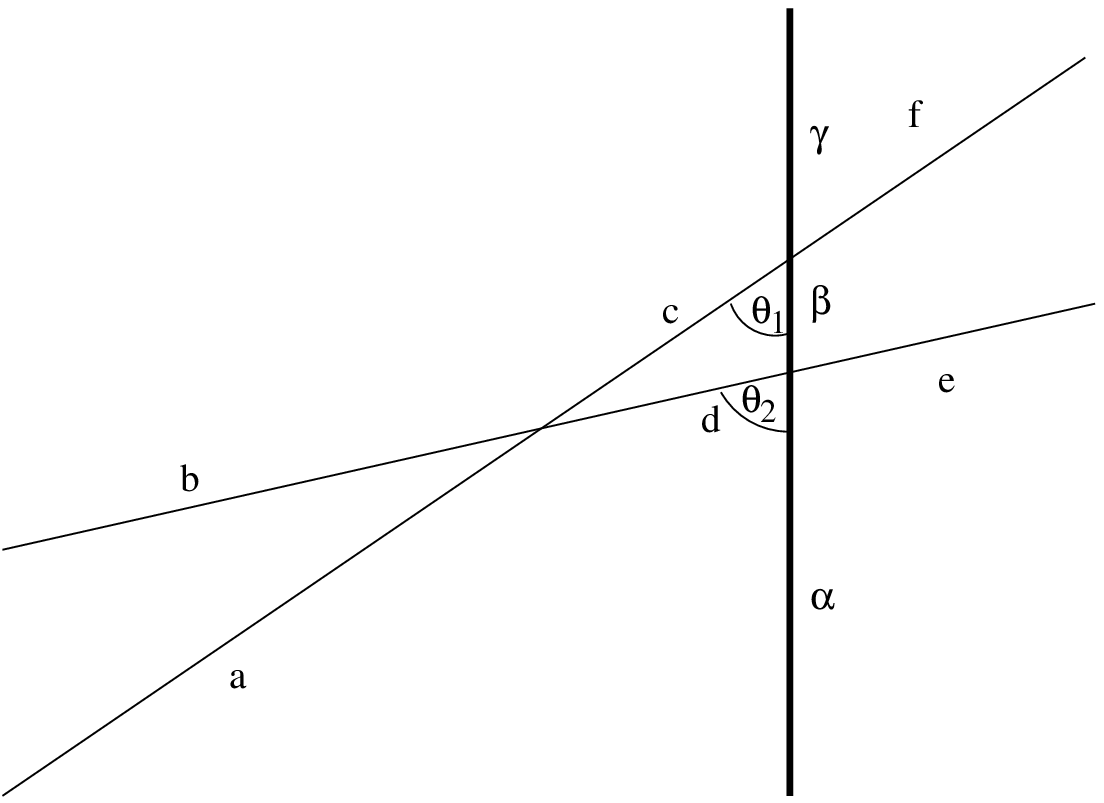}\end{center}

\begin{equation}
{\TL}_{\alpha a}^{\beta d}(\theta_{1}){\TL}_{\beta b}^{\gamma c}(\theta_{2}){\SR}_{dc}^{fe}(\theta_{2}-\theta_{1})\qquad=\qquad{\SL}_{ab}^{cd}(\theta_{2}-\theta_{1}){\TL}_{\alpha d}^{\beta e}(\theta_{2}){\TL}_{\beta c}^{\gamma f}(\theta_{1})\,\,.\label{te}\end{equation}
 We can now act on the defect system with the operator $\mathcal{P}_{\Psi}$.
Diagrammatically this flips the right side of the defect Yang-Baxter
equation onto the left of the boundary. We see this is a special case
of the boundary Yang-Baxter equation where both reflections cause
a change in the species of the reflecting particles. This corresponds
to how, as we already know, $\mathcal{P}_{\Psi}$ maps the transmission
matrix $\TL$ into the $\mathcal{R}_{\Phi\Psi}$ element of the $R$-matrix.
Different reflection transmission equations correspond to other special
cases of the boundary Yang-Baxter equation. The boundary Yang-Baxter
diagram that is equivalent to \eqref{te} and the corresponding equation
is 

\begin{center} \includegraphics[
height=4cm,
keepaspectratio]{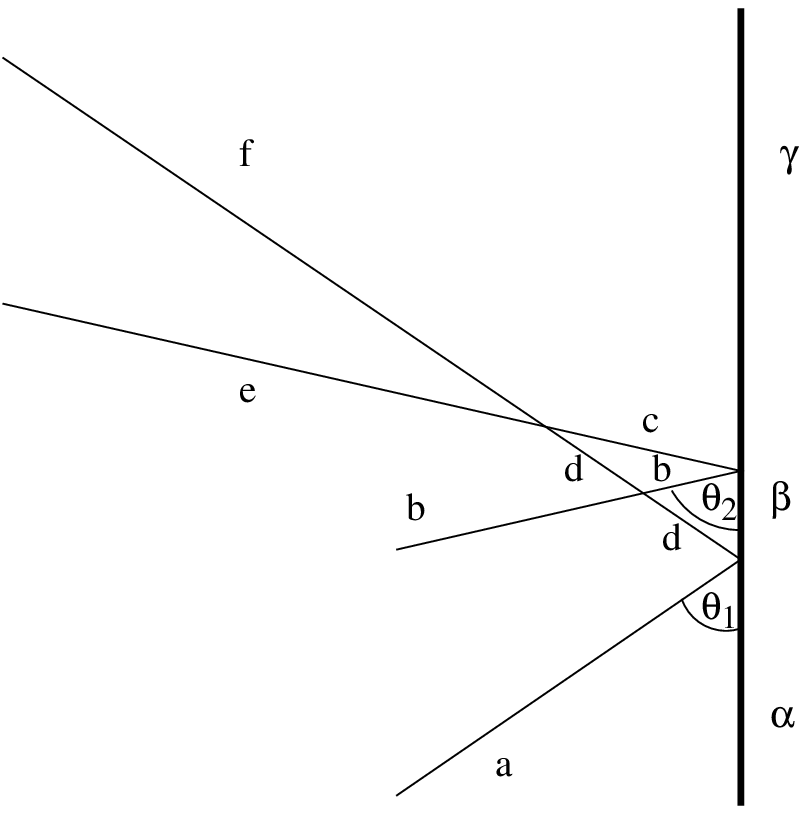} \qquad\qquad\qquad
\includegraphics[
height=4cm,
keepaspectratio]{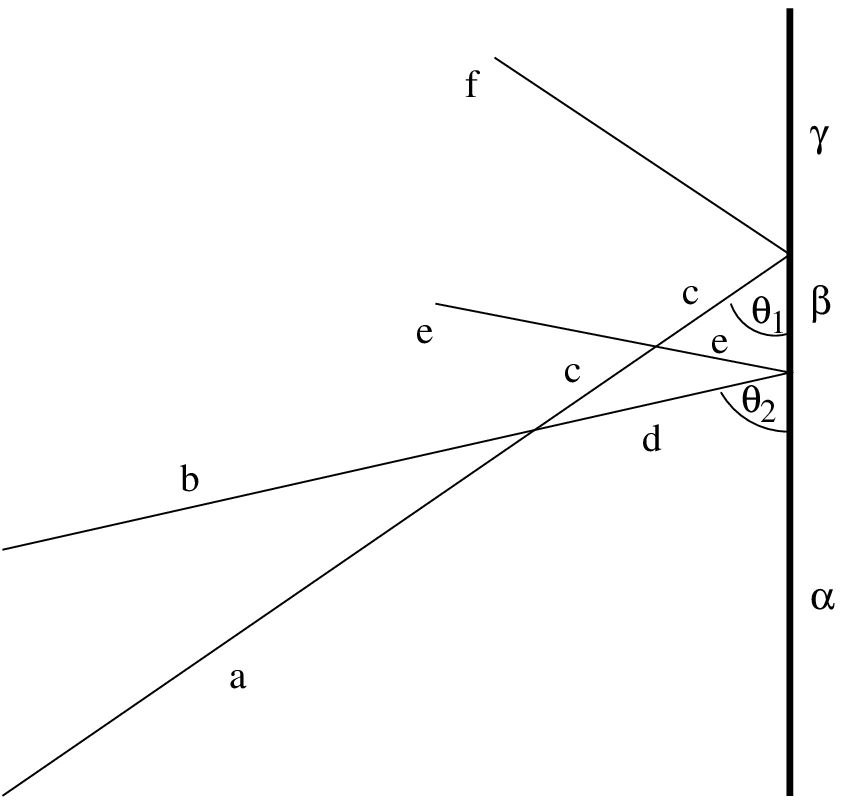}\end{center}

\begin{align}
{\mathcal{R}_{\Phi\Psi}}_{\alpha a}^{\beta d}(\theta_{1}){\mathcal{S}_{\Phi\Psi}}_{bd}^{bd}(\theta_{1}+\theta_{2}) & {\mathcal{R}_{\Phi\Psi}}_{\beta b}^{\gamma c}(\theta_{2}){\mathcal{S}_{\Psi\Psi}}_{dc}^{fe}(\theta_{2}-\theta_{1})\nonumber \\
 & ={\mathcal{S}_{\Phi\Phi}}_{ab}^{cd}(\theta_{2}-\theta_{1}){\mathcal{R}_{\Phi\Psi}}_{\alpha d}^{\beta e}(\theta_{2}){\mathcal{S}_{\Psi\Phi}}_{ce}^{ce}(\theta_{1}+\theta_{2}){\mathcal{R}_{\Phi\Psi}}_{\beta c}^{\gamma f}(\theta_{1})\,\,,\label{bYBE}\end{align}
 where the `\emph{in}' states, labeled $a$ and $b$, belong to the
field $\Phi$ and the `\emph{out}' states $e$ and $f$ belong to
$\Psi$. 

The reflection and transmission amplitudes of the defect satisfy the
unitarity equations \begin{align}
{\RL}_{\alpha a}^{\beta b}(\theta){\RL}_{\beta b}^{\gamma c}(-\theta)+{\TL}_{\alpha a}^{\beta b}(\theta){\TR}_{\beta b}^{\gamma c}(-\theta)= & \,\,\delta_{a}^{c}\delta_{\alpha}^{\gamma}\,\,,\nonumber \\
{\TR}_{a\alpha}^{b\beta}(\theta){\TL}_{b\beta}^{c\gamma}(-\theta)+{\RR}_{a\alpha}^{b\beta}(\theta){\RR}_{b\beta}^{c\gamma}(-\theta)= & \,\,\delta_{a}^{c}\delta_{\alpha}^{\gamma}\,\,,\nonumber \\
{\RL}_{\alpha a}^{\beta b}(\theta){\TL}_{b\beta}^{c\gamma}(-\theta)+{\TL}_{\alpha a}^{\beta b}(\theta){\RR}_{\beta b}^{\gamma c}(-\theta)= & \,\,0\,\,,\nonumber \\
{\TR}_{\alpha a}^{\beta b}(\theta){\RL}_{\beta b}^{\gamma c}(-\theta)+{\RR}_{\alpha a}^{\beta b}(\theta){\TR}_{\beta b}^{\gamma c}(-\theta)= & \,\,0\,\,.\label{dunit}\end{align}
 The $\mathcal{R}$-matrix has to satisfy the unitarity equation ${\mathcal{R}_{a\alpha}^{b\beta}(\theta)\mathcal{R}_{b\beta}^{c\gamma}(-\theta)=\delta_{a}^{c}\delta_{\alpha}^{\gamma}}$,
which, in terms of the defect variables, reads as equation \eqref{dunit}.
The $\mathcal{R}$-matrix must also satisfy the boundary crossing
unitarity equation \cite{GZ}, ${\mathcal{R}_{\alpha a}^{\beta b}(\theta)=\mathcal{S}_{ad}^{cb}(2\theta)\mathcal{R}_{\alpha\bar{d}}^{\beta\bar{c}}(i\pi-\theta)}$,
in terms of the defect this gives \begin{align}
 & {{\RL}_{\alpha a}^{\beta b}(\theta)={\SL}_{ad}^{cb}(2\theta){\RL}_{\alpha\bar{d}}^{\beta\bar{c}}(i\pi-\theta)\,\,,} & {{\TL}_{\alpha a}^{\beta b}(\theta)={\TR}_{\alpha\bar{b}}^{\beta\bar{a}}(i\pi-\theta)\,\,,}\nonumber \\
 & {{\RR}_{\alpha a}^{\beta b}(\theta)={\SR}_{ad}^{cb}(2\theta){\RR}_{\alpha\bar{d}}^{\beta\bar{c}}(i\pi-\theta)\,\,,} & {{\TR}_{\alpha a}^{\beta b}(\theta)={\TL}_{\alpha\bar{b}}^{\beta\bar{a}}(i\pi-\theta)\,\,,}\label{dxunit}\end{align}
 observe, however, that these equations are \emph{derived} and not
just conjectured as in \cite{DMS1,DMS2}. Now consider the consequences
of these equations. Since they are the same as in \cite{DMS1,DMS2,CAFG}
we have analogous consequences; for rapidity dependent bulk $S$-matrices,
with no degrees of freedom at the defect, there is no simultaneous
transmission and reflection. Equivalently, for the boundary system
described earlier, reflections are either species changing or species
preserving.

Let us now derive the defect operator using the defect-boundary correspondence for 
the model containing one self-conjugated particle as defined in \eqref{defectL}. 
Consider first the boundary system as described in \eqref{boundaryL2}. 
If we change the roles of space and time the boundary becomes an initial state
created by a boundary operator $B$ from the vacuum $\vert B\rangle=B\vert 0\rangle$.
This operator, using the Zamolodchikov-Faddeev creation annihilation operators, 
  is given by \cite{GZ} as follows: 
\begin{eqnarray*}
B & = & \exp\left\{ \int_{0}^{\infty}\frac{d\theta}{2\pi}\left(R_{\Phi \Phi}\left(\frac{i\pi}{2}-\theta\right)A_{\PhiL}^{+}(-\theta)A_{\PhiL}^{+}(\theta)+R_{\Phi \Psi}\left(\frac{i\pi}{2}-\theta\right)A_{\PhiL}^{+}(-\theta)A_{\PsiL}^{+}(\theta)\right.\right.\\
 &  & \left.\left.\qquad\qquad\qquad+R_{\Psi \Phi}\left(\frac{i\pi}{2}-\theta\right)A_{\PsiL}^{+}(-\theta)A_{\PhiL}^{+}(\theta)\right)+R_{\Psi \Psi}\left(\frac{i\pi}{2}-\theta\right)A_{\PsiL}^{+}(-\theta)A_{\Psi}^{+}(\theta)\right\} \end{eqnarray*}
 Now if we unfold the system the time boundary becomes a defect line,
where the defect operator is inserted. Using the relation \eqref{RRT} and making
the change $A_{\PsiL}^{+}(\theta)\rightarrow A_{\PsiR}(-\theta)$
we get the defect operator 
\begin{eqnarray*}
D & = & \exp\left\{ \int_{0}^{\infty}\frac{d\theta}{2\pi}\left(R_{\mathbb{L}}\left(\frac{i\pi}{2}-\theta\right)A_{\PhiL}^{+}(-\theta)A_{\PhiL}^{+}(\theta)+T_{\mathbb{L}}\left(\frac{i\pi}{2}-\theta\right)A_{\PhiL}^{+}(-\theta)A_{\PsiR}(-\theta)\right.\right.\nonumber \\
 &  & \left.\left.\qquad\qquad\quad+T_{\mathbb{R}}\left(\frac{i\pi}{2}-\theta\right)A_{\PsiR}(\theta)A_{\PhiL}^{+}(\theta)\right)+R_{\mathbb{R}}\left(\frac{i\pi}{2}-\theta\right)A_{\PsiR}(\theta)A_{\PsiR}(-\theta)\right\} \end{eqnarray*}
A similar operator was proposed in \cite{defop} but without any derivation.
Our \emph{derivation} confirms their result. Observe that since the $\Phi$ and $\Psi$
particles do not interact $A_{\PsiR}(\theta_1)A^{+}_{\PhiL}(\theta_2)=
A^{+}_{\PhiL}(\theta_2)A_{\PsiR}(\theta_1)$ and no normal ordering is needed. 
We note also that as a consequence of equation \eqref{dxunit} $ T_{\mathbb{L}}\left(\frac{i\pi}{2}+\theta\right)=
T_{\mathbb{R}}\left(\frac{i\pi}{2}-\theta\right)$ so the defect operator in the 
purely transmitting case simplifies to: 
\begin{equation} 
 D=\exp\left\{ \int_{-\infty}^{\infty}\frac{d\theta}{2\pi}\left(T_{\mathbb{R}}\left(\frac{i\pi}{2}-\theta\right)A_{\PsiR}(\theta)A_{\PhiL}^{+}(\theta)\right)\right\} 
\label{defop}
\end{equation}
\section{Purely transmitting defects}

\smallskip
As integrable defect theories with nontrivial S-matrix can only have
purely transmitting or reflecting defects we can consider
how to find solutions in these two cases separately. If a defect is
purely reflecting then we can solve the $\TheoL$ and $\TheoR$ parts
independently, these equations are not coupled contrary to what is
claimed in \cite{CAFG}. Thus let us concentrate on the purely transmitting
defects, in the boundary language this means that the refection matrix
is purely off diagonal. One can obtain nontrivial solutions of the
boundary Yang-Baxter equation by the fusion method as follows. 

Suppose we have found a family of solutions, $\mathcal{R}_{c\alpha}^{e\beta}(\theta),\left\{ \alpha\right\} ,\left\{ \beta\right\} $
of the boundary Yang-Baxter \eqref{bYBE}, unitarity and boundary
crossing unitarity equations. Now we can produce another solution,
for an excited boundary, via the boundary bootstrap equation \cite{GZ}

\begin{center} \includegraphics[
height=4cm,
keepaspectratio]{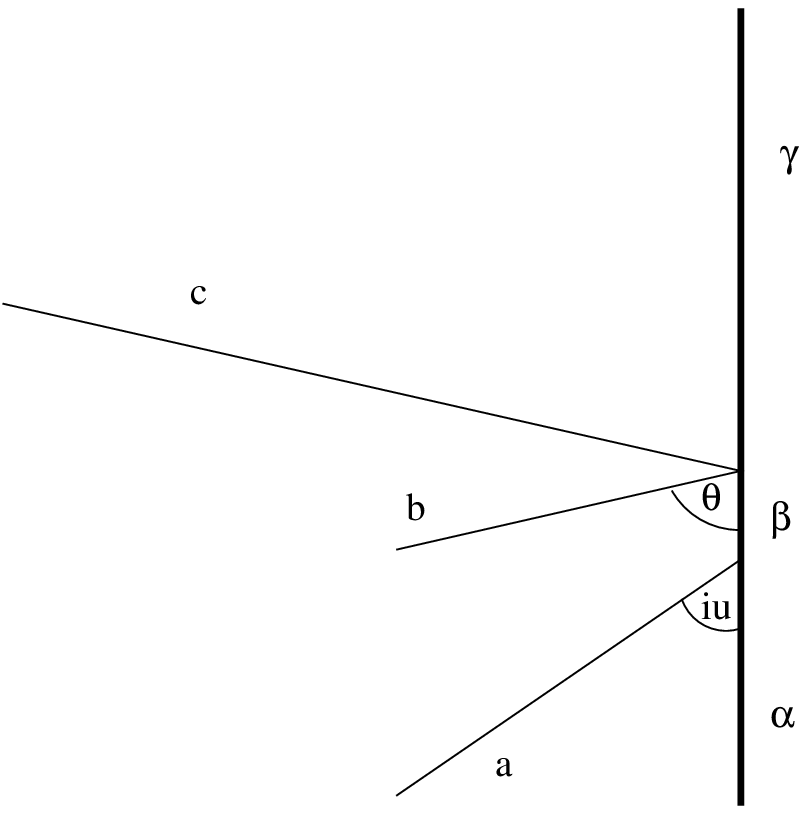} \qquad\qquad\qquad
\includegraphics[
height=4cm,
keepaspectratio]{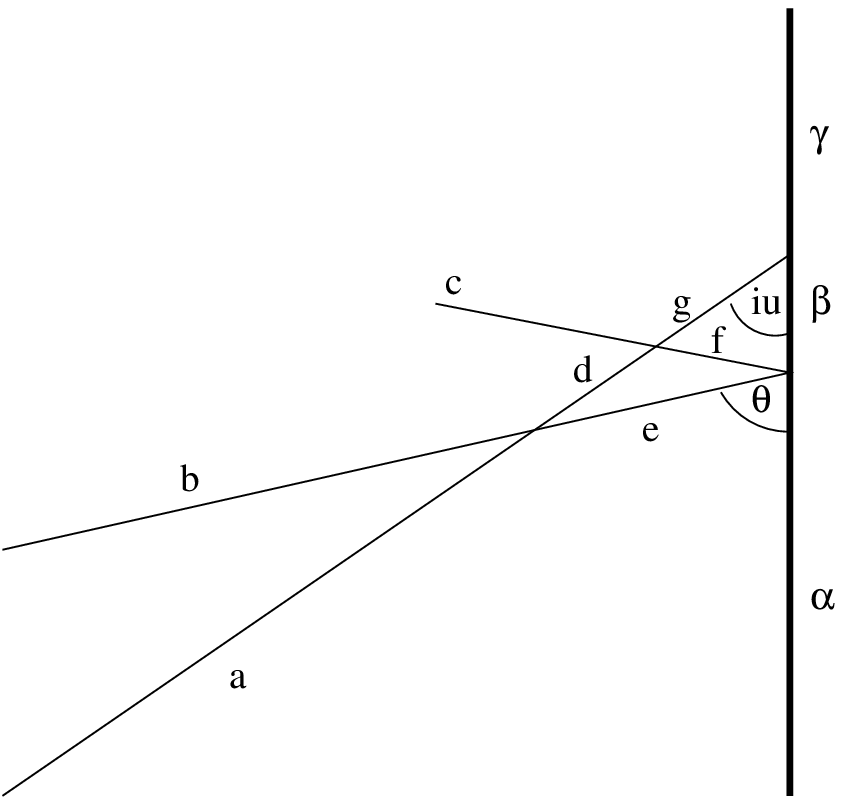}\end{center}

\begin{equation}
\qquad\qquad\qquad g_{a\alpha}^{\beta}\mathcal{R}_{b\beta}^{c\gamma}(\theta)\qquad=\qquad g_{g\beta}^{\gamma}\mathcal{S}_{ab}^{de}(\theta-iu)\mathcal{R}_{e\alpha}^{f\beta}(\theta)\mathcal{S}_{fd}^{eg}(\theta+iu)\,\,.\label{bootstrap}\end{equation}
 This new solution, $\mathcal{R}_{b\beta}^{c\gamma}(\theta)$, automatically
satisfies the required equations. The proof of the boundary Yang-Baxter
equation or factorization can be shown first by `shifting' the trajectories
from the excited boundary to the ground state boundary, then using
the proof on this boundary and shifting back to the excited boundary.
The unitarity and boundary crossing unitarity is a consequence of
the similar equations for the bulk $S$-matrix. If the original $R$-matrix
does not have any pole in the physical strip we can always add a CDD
factor to ensure its existence. By applying the folding method to
transmitting defect systems as described in this letter we can apply
this method to find non-trivial solutions of defect systems. 

For example let us assume that $\mathcal{S}_{\Phi\Phi}$ and $\mathcal{S}_{\Psi\Psi}$
are parity transforms of each other ${\mathcal{S}_{\Phi\Phi}}_{ab}^{cd}(\theta)={\mathcal{S}_{\Psi\Psi}}_{ba}^{dc}(\theta)$.
In this case the solution $\mathcal{R}_{\Phi\Psi}=\mathcal{R}_{\Psi\Phi}=1$,
which corresponds to trivial transmission, satisfies all the required
equations. By adding a CDD factor to this solution without spoiling
its nice properties we can ensure the existence of a pole at $\theta=iu$,
$0<u<\pi/2$. Let us label this unexcited boundary solution by $0=\alpha$.
Now if a particle with label $a$ is bound to the boundary the excited
boundary state will have label $a=\beta$. A particle of type $b$
can reflect from the boundary to particle $c$ by changing the boundary
state from $a$ to $\gamma=d$ as shown on the previous figure. By
using \eqref{bootstrap} we can find an expression for the reflection
matrix for this excited boundary. The $\mathcal{R}$-matrix for the
excited boundary now depends upon the species of reflecting particle,
\begin{align}
 & {{\mathcal{R}_{\Phi\Psi}}_{ba}^{cd}(\theta)={T_{\mathbb{L}}}_{ba}^{cd}={\mathcal{S}_{\Phi\Phi}}_{ab}^{cd}(\theta-iu)\,\,;}
& {{\mathcal{R}_{\Psi\Phi}}_{ba}^{cd}(\theta)={T_{\mathbb{R}}}_{ba}^{cd}={\mathcal{S}_{\Psi\Psi}}_{ba}^{dc}(\theta+iu)\,\,.}\label{exRmat}\end{align}
 Clearly they satisfy (\ref{te}), since the bulk $S$-matrix satisfies
the bulk Yang-Baxter equation. They also satisfy the unitarity and
cross unitarity relations since it reduces to the analogous equations
of the bulk $S$-matrix. Observe that equation (\ref{te}) is similar
to the Lax equations of integrable lattice models and so the bulk
S matrix always solves it. If $u=0$ the solution describes a standing
particle. Bootstrapping further on this excited defect we can bind
either different type, or more particles to the defect. It is similar
to how one can find higher spin solutions of the Lax equation in lattice
models. 

We note that another special solution for purely transmitting impurity
is found by Konik and LeClair, for the sine-Gordon model using the
perturbed CFT approach \cite{KL}. 

Purely transmitting defects are of particular interest in the study of
boundary systems as they may be
used to construct new solutions to the reflection equation. 
If there exists a ${\mathcal{T}_\mathbb{L}}_{a\alpha}^{b\beta}(\theta)$
and ${\mathcal{T}_\mathbb{R}}_{a\alpha}^{b\beta}(\theta)$ which 
satisfies the transmission equation \eqref{te}, unitarity \eqref{dunit}
and crossing unitarity \eqref{dxunit}
for some $S$-matrix, $\mathcal{S}_{ab}^{cd}(\theta)$, and a
${\mathcal{R}_0}_{a\alpha}^{b\beta}(\theta)$ that satisfies
the reflection equation \eqref{bYBE} for the same $S$-matrix then the
matrix 
\begin{equation}
{\mathcal{R}_1}_{a\alpha}^{b\beta}(\theta)=
{\mathcal{T}_\mathbb{L}}_{a\alpha_1}^{c\gamma_1}(\theta)
{\mathcal{R}_0}_{c\alpha_2}^{d\beta_2}(\theta)
{\mathcal{T}_\mathbb{R}}_{d\gamma_1}^{b\beta_1}(-\theta)\ ,
\end{equation}
where $\alpha=\alpha_1\otimes\alpha_2$ and
$\beta=\beta_1\otimes\beta_2$,  
also satisfies the reflection equation \cite{S2}. This is diagrammatically
represented below.

\begin{center} \includegraphics[
height=4cm,
keepaspectratio]{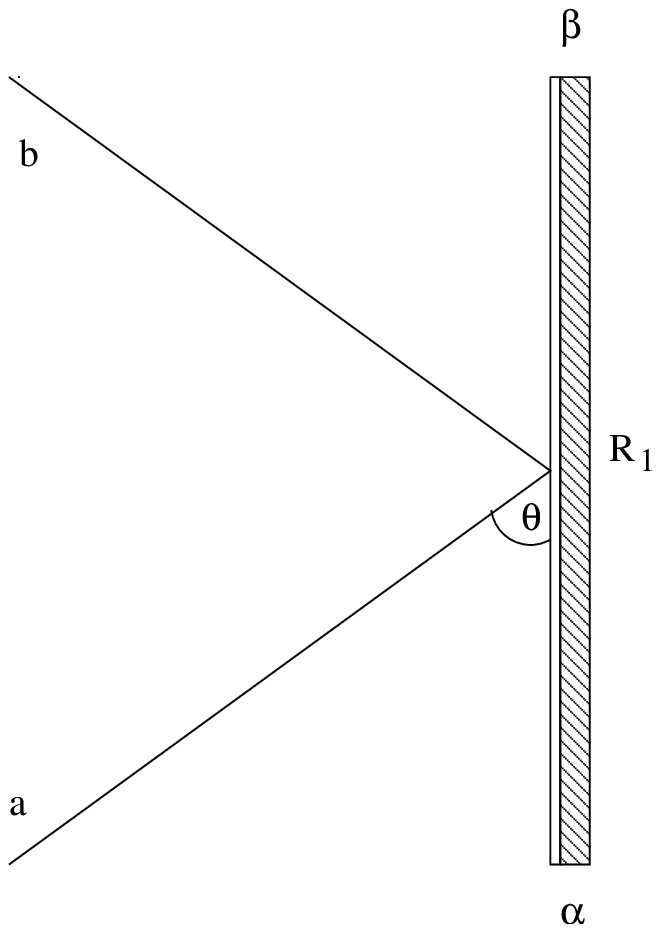} \qquad\qquad\qquad
\includegraphics[
height=4cm,
keepaspectratio]{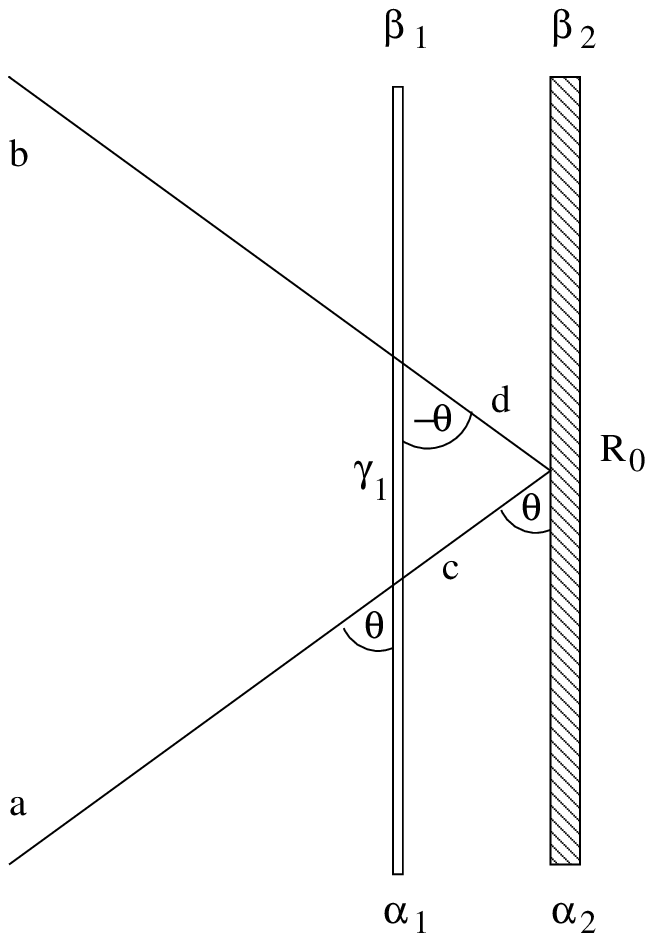}\end{center}

The $R$-matrices obtained in this way are called dressed $R$-matrices in
\cite{Zab}.
If $\mathcal{R}_1$ is block diagonal in the $\alpha$, $\beta$ basis then
the $R$-matrix can be projected into one of the irreducible blocks, the
defect is then said to be fused to the boundary.
Solutions developed in this manner, using \eqref{exRmat}, are equivalent to
performing the boundary bootstrap procedure.

\section{Application: defect TBA}

Let us now consider one application of the boundary-defect correspondence
developed in this paper; suppose we would like to calculate the groundstate
energy of an integrable system, containing one particle type with mass $m$, and 
bulk scattering matrix $S$,  in a finite geometry. The system is
defined on an interval of size $L$ with integrable boundary conditions
at both ends, and $n$ integrable purely transmitting defects with
the $i^{th}$ defect localized at $x_{i}$ as shown in the following
figure

\bigskip{}
\begin{center}\includegraphics[%
  width=10cm]{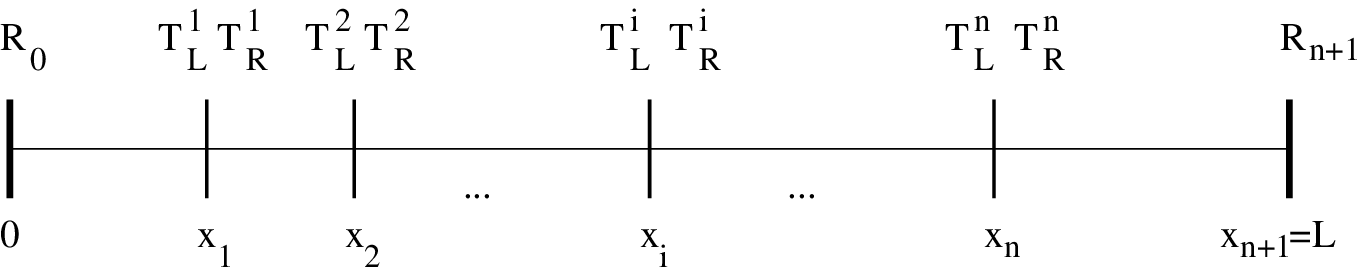}\end{center}
\medskip{}

\bigskip{} The usual trick is to compactify the time-like direction
on to a circle of size $R$ and compute the partition function of
the compactified system, $Z(L,R)$, which is related to its energy,
$E$, as \[
Z(L,R)=Tr(e^{-RE(L)})\quad.\]
 The groundstate energy of the original system, $E_{0}$, can be found
by taking the limit as $R$ goes to infinity \[
E_{0}(L)=-\lim_{R\rightarrow\infty}\frac{1}{R}Z(L,R)\qquad.\]
 We now exchange the role of time and space and calculate the partition
function in the following geometry:

\bigskip{}
\begin{center}\includegraphics[%
  height=7cm]{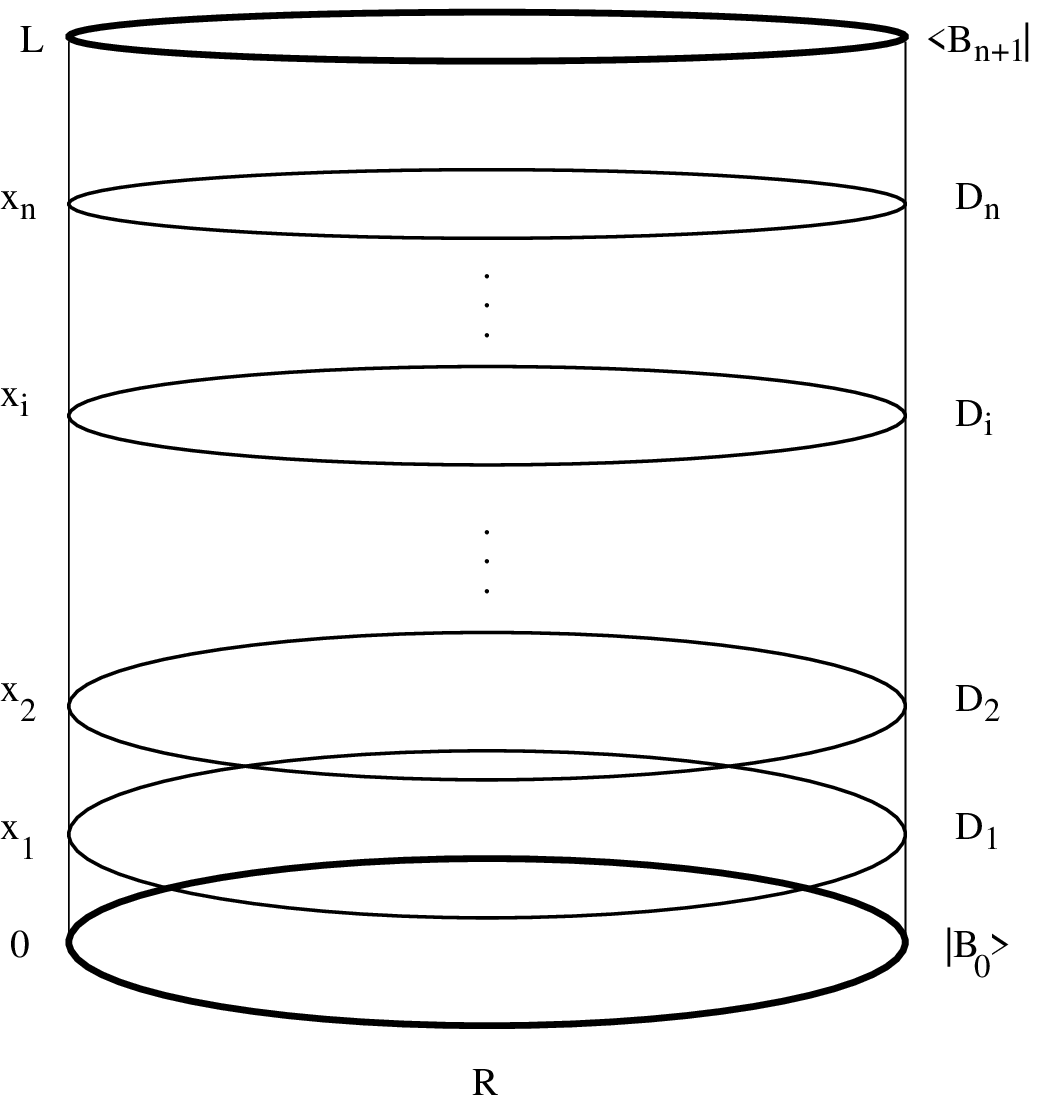}\end{center}
\medskip{}

\bigskip{} The Hilbert spaces used are now the periodic Hilbert spaces,
the boundaries become initial and final states, while the defects
show up as operators which have to be inserted into the matrix element
\begin{eqnarray}
Z(L,R) & = & \langle B_{n+1}|e^{-(L-x_{n})H^{n+1}(R)}D_{n}e^{-(x_{n}-x_{n-1})H^{n}(R)}D_{n-1}\dots e^{-(x_{i}-x_{i-1})H^{i}(R)}D_{i}\dots\nonumber \\
 &  & e^{-(x_{2}-x_{1})H^{2}(R)}D_{1}e^{-x_{1}H^{1}(R)}|B_{0}\rangle\qquad.\label{eq:dpart}\end{eqnarray}
$H^{i}(R)$ is the Hamiltonian of the system between $x_{i-1}$ and
$x_{i}$ with periodic boundary condition. In order to perform the
calculation of the partition function we have to use the form of
the defect operators \eqref{defop}.

We calculate the partition function by introducing the resolution
of the identity into (\ref{eq:dpart}). Using the $N$-particle matrix
element \[
\langle A_{i+1}^{+}(\theta_{1})A_{i+1}^{+}(\theta_{2})\dots A_{i+1}^{+}(\theta_{N})|D_{i}|A_{i}^{+}(\theta_{1}^{\prime})A_{i}^{+}(\theta_{2}^{\prime})\dots A_{i}^{+}(\theta_{N}^{\prime})\rangle=\prod_{j=1}^{N}T_{\mathbb{R}}^{i}\left(\frac{i\pi}{2}-\theta_j\right)2\pi\delta(\theta_{j}-\theta_{j}^{\prime})\]
 we obtain \[
Z(L,R)=\sum_{|N\rangle\in\mathcal{H}_{N}}\frac{\langle B_{n+1}|N\rangle\langle N|B_{0}\rangle}{\langle N|N\rangle}e^{-L\sum_{j=1}^{N}\left(m\cosh(\theta_{j})+\sum_{k=1}^{n}\log\left(T_{\mathbb{R}}^{k}\left(\frac{i\pi}{2}-\theta_{j}\right)\right)\right)}\]
 As we can see the partition function does not depend on the actual
positions of the defects, which is a consequence of the integrability
of the system. The further calculations follow exactly the one in
\cite{LMSS} with the result for the ground state energy \begin{equation*}
E_{0}(L) = -\frac{m}{4\pi}\int_{-\infty}^{-\infty}\cosh(\theta)\log(1+\lambda(\theta)e^{-\epsilon(\theta)})d\theta
\end{equation*}
where $\epsilon(\theta)$ satisfies the following integral equation
\begin{equation*}
\epsilon(\theta) = 2mL\cosh(\theta)+\int_{-\infty}^{-\infty}\Phi(\theta-\theta^{\prime})\log(1+\lambda(\theta^{\prime})e^{-\epsilon(\theta^{\prime})})d\theta^{\prime}\end{equation*}
 where $\Phi(\theta)=-\frac{1}{2\pi i}\frac{d}{d\theta}S(\theta)$
The only difference compared to \cite{LMSS} is that now \[
\lambda(\theta)=R_{0}\left(\frac{i\pi}{2}+\theta\right)R_{n+1}\left(\frac{i\pi}{2}-\theta\right)\prod_{j}T_{\mathbb{L}}^{j}\left(\frac{i\pi}{2}+\theta\right)T_{\mathbb{R}}^{j}\left(\frac{i\pi}{2}-\theta\right)\qquad.\]

 Note that if all the defects have trivial transition matrices we
reproduce the formula of \cite{LMSS}. We can interpret the result
as follows; since nothing depends on the position of the defects we
can move them all to either of the boundaries. Now we do not have
any defect, but the boundary is dressed, and what we can see in the
final result is nothing but the dressed reflection factors.

\section{Conclusion}
\smallskip
Using the folding trick we have shown that defect theories are equivalent to certain boundary theories. 
As a consequence  we can take results from boundary
field theory, such as reduction formulas, Coleman-Thun mechanism,
Cutcosky rules. 

Applying this folding trick to integrable systems the 
defect crossing unitarity and defect operator were derived, 
and the fusion method was used for finding solutions  to general 
impurity systems. 
The defect operator is used to calculate the groundstate energy of an integrable 
system defined on the interval with defects. Since the result does not 
depend on the location of the defects, defects can be fused to any 
of the boundaries obtaining a system without defects but with dressed 
reflection matrices. 

This idea can be used in a theory with non-diagonal 
boundary scattering as follows. Suppose we can write the non-diagonal reflection
factor as a diagonal one fused with a non-diagonal defect. Now moving this 
non-diagonal defect to the other boundary and fusing with it the fused 
reflection factor may be diagonal helping in the solution of the model.  

\subsection*{Acknowledgments}

The authors would like to thank L\'{a}szl\'{o} Palla, G\'{a}bor
Tak\'{a}cs, Gustav Delius, Atsushi Higuchi and Peter Bowcock for interesting and helpful discussions.
We also thank Eric Ragoucy and Mihail Mintchev for calling our attention
to the importance of the assumption of relativistic invariance. ZB is supported by
a Bolyai J\'{a}nos Research Fellowship and FKFP 0043/2001, OTKA T043582,
OTKA T037674. AG is supported by a PPARC studentship. This paper
is a result of discussions undertaken during a EUCLID, HPRN-CT-2002-00325,
funded visit.

\end{document}